# Post-Quantum Secure Federated DeFi for Inclusive Banking

S. Sachan[1], D. Fickett[2], R. Buchinger[3], and T. Miller[3]

[1]S. Sachan is with the AI FinTech Group, University of Liverpool, L69 3BX, UK (Swati.Sachan@liverpool.ac.uk ).
[2]D. Fickett is with the RVA Works and University of Richmond, Richmond, VA 23173, USA (dfickett@richmond.edu ).
[3]R. Buchinger and T. Miller is with Chain Crunch Labs based in Finsbury, London, UK and Kendall Square, Cambridge, Massachusetts, MA 02139, USA (Richard.Buchinger@chaincrunchlabs.com , Theo.Miller@chaincrunchlabs.com ).

***Abstract*— Recent advances in error-corrected qubits have accelerated the timeline for practical quantum computing. It poses a threat to cryptographic primitives used to secure financial systems, government infrastructure, communication networks, and DeFi (Decentralized Finance) ecosystems. This paper introduces a post-quantum secure federated DeFi framework that enables inter-bank collaboration to improve the inclusivity of individuals underserved by local lenders due to limited financial histories. Multiple banks contribute encrypted information batches to a virtual server, where lattice-based Fully Homomorphic Encryption (FHE) enables end-to-end homomorphic computation. The server fuses local data-driven probabilistic assessments, expert beliefs, and verifiable evidence generated by the NASA-IBM Prithvi Geospatial Foundation Model (GFM), in encrypted format. Decentralized technologies are employed to ensure tamper-proof evidence and auditable accountability for all encrypted data exchanges between institutions and the server. The framework is tested on agricultural lending decisions for rural borrowers in Virginia.**

## I. Post-Quantum Security and DeFi for Secure AI

Recent breakthroughs in error-corrected qubits achieved by leading technology companies have compressed expectations for scalable quantum computing systems from decades to just a few years [1]. Logical qubits are engineered within physical quantum hardware to detect and correct computational errors and maintain quantum coherence to preserve reliable information over extended operations.

The transformative potential of quantum computing presents a two-sided paradigm: it promises breakthroughs in fields such as drug discovery, materials science, and optimization problems, but also poses a threat to the foundations of modern cybersecurity. The security of global financial, governmental, and communication infrastructures currently relies on the computational hardness of mathematical problems such as integer factorization, which underpins the RSA algorithm, and discrete logarithms, which form the basis of elliptic curve cryptography and finite-field Diffie–Hellman key exchange. Additionally, the foundational security mechanisms used in Decentralized Finance (DeFi) ecosystems are susceptible to future quantum attacks.

The vulnerability becomes even more pronounced as decentralized technologies increasingly integrate AI-driven analytics for advanced risk modelling, decision automation, and collective intelligence in financial operations. A successful quantum breach could not only expose sensitive financial and transactional information but also compromise the integrity of AI models trained on sensitive historical datasets. The consortium-based financial intelligence application, such as the detection of money laundering, relies on classical cryptographic primitives for Secure Multi-Party Computation (SMPC) and federated data sharing.

Quantum algorithms such as Shor's algorithm could efficiently break public-key cryptography in polynomial time, and Grover's algorithm could weaken symmetric encryption and hash-based security [2]. Quantum-enabled cryptanalysis exacerbates "harvest now, decrypt later" threats, where adversaries collect and store encrypted information today for decryption once quantum capabilities mature [3]. It points to the urgent need for a global migration toward post-quantum cryptography (PQC) to maintain trust, security, and interoperability in AI-driven decentralized financial systems.

Lattice-based cryptography is widely regarded as one of the most promising candidates for PQC due to its strong security guarantees against both classical and quantum attacks. Its security is reducible to the computational intractability of core lattice problems on the Shortest Vector Problem (SVP), Closest Vector Problem (CVP), and Learning with Errors (LWE) [4]. It is the core of advanced cryptographic primitives such as Fully Homomorphic Encryption (FHE), Zero-Knowledge Proofs (ZKPs), and SMPC, which are essential for privacy-preserving AI techniques and confidential computation in DeFi systems. A lattice $\mathcal{L} \subseteq \mathbb{R}^n$ is a discrete additive subgroup formed by all integer linear combinations of a basis consisting of $n$ linearly independent vectors $b_1, \ldots, b_n \in \mathbb{R}^n$:

$$\mathcal{L} = \left\{ \sum_{j=1}^{n} c_j \, b_j \middle| c_j \in \mathbb{Z} \right\} \quad (1)$$

The coefficients $c_j$ are constrained to the integers $\mathbb{Z}$ to ensure the lattice of $n$-dimensional Euclidean space over the real numbers $\mathbb{R}^n$prevents dense filling to maintain minimum distances between points. It is a key property of the hardness of SVP (finding the shortest non-zero vector) and CVP (finding the closest lattice point to a given target). Each basis vector $b_j$ spans in a unique direction in the space; linear dependence among them would create a degenerate lattice of reduced dimensionality, which compromises security by weakening the hardness [5]. The LWE problem is a famous lattice-based cryptosystem based on the hardness of finding random noisy linear equations of a secret vector $\mathscr{s} \in \mathbb{Z}_q^n$, hidden in equations of the form $a_j \cdot \mathscr{s} + \varepsilon_j \equiv b_j (mod\ q)$, where $\varepsilon_j$ are small Gaussian errors [6].

Variants such as Ring Learning with Errors (RLWE) [7] and Torus Learning with Errors (TRLWE) [8] extend LWE to polynomial rings for efficient FHE schemes used for end-to-end encrypted computation. RLWE operates over the cyclotomic ring $\mathbb{Z}_{q,N}[X] = \frac{\mathbb{Z}_q[X]}{(X^N+1)}$, where $q$ is a large prime modulus, $N$ is a power of two, and $X^N + 1$ is a cyclotomic polynomial. This ring structure enhances computational efficiency and enables compact key representation compared to standard LWE. TRLWE generalizes RLWE to the torus polynomial ring $\mathbb{T}_N[N] = \frac{(\mathbb{R}/\mathbb{Z})[X]}{(X^N+1)}$, where coefficients lie in the real torus modulo 1. It allows continuous error distributions and supports high-precision arithmetic in advanced FHE frameworks such as TFHE (Fast Fully Homomorphic Encryption over the Torus).

Lattice-based FHE schemes include Cheon-Kim-Kim-Song (CKKS), Brakerski-Fan-Vercauteren (BFV), and TFHE [9]. The CKKS scheme is an RLWE-based FHE system optimized for approximate arithmetic on real and complex numbers, making it ideal for privacy-preserving machine learning and financial analytics. TFHE is built on TRLWE; it excels at fast bootstrapping for the efficient evaluation of arbitrary Boolean and arithmetic circuits on encrypted data.

This paper presents a post-quantum secure DeFi framework for inter-bank collaboration to enhance the inclusivity of individuals underserved by local banks due to conservative lending practices and limited credit histories. As illustrated in Figure 1, the framework aggregates encrypted batches of information comprising three evidence sources: probabilistic decisions derived from local bank data, expert belief assessments, and verifiable inferences generated by Geospatial Foundation Model (GFMs), such as land-fertility evaluation from satellite imagery in the rural agricultural lending case study presented in this paper. Human–AI belief and probability masses are fused using Dempster Shafer Theory (DST) of uncertainty at a virtual server, entirely in encrypted form by FHE schemes based on RLWE and TRLWE. Additionally, decentralized technologies are employed to ensure tamper-proof evidence and auditable accountability for all encrypted data exchanges between financial institutions and the server.

The rest of the paper is organized as follows. Section II reviews related work on privacy-preserving AI enabled by lattice-based cryptography. Section III introduces the proposed framework for post-quantum decentralized multi-bank collaboration. Section IV presents the case study of an rural agricultural lending. The paper is concluded in Section V.

## II. Related Work

Existing studies on privacy-preserving AI employing cryptographic techniques, particularly FHE, have predominantly focused on SMPC and federated learning (FL) frameworks for encrypting and aggregating local model parameter updates in Internet of Things (IoT) and edge computing environments [10][11]. Recently, Zama has advanced the practical deployment of FHE through fhEVM, an extension of the Ethereum Virtual Machine (EVM) that enables encrypted state management and computation within smart contracts [12]. The TFHE scheme by Zama supports privacy-preserving decentralized applications such as confidential banking data aggregation with AI-driven decisions, decentralized exchanges (DEXs), and auctions by enabling high-speed computation directly on encrypted data without revealing plaintext information [13].

Privacy-preserving lending in DeFi enables users to demonstrate over-collateralization without disclosing exact collateral or debt amounts, thereby safeguarding financial privacy and transactional confidentiality. Recent research integrated FHE with ZKPs to achieve verifiable yet private credit validation [14]. Furthermore, the convergence of AI with decentralized technologies such as Blockchain and the InterPlanetary File System (IPFS) for financial decision-making has been proposed to ensure traceable accountability, secure auditability, and responsible AI governance. These systems monitor collaborative human-AI activities and the quality of GFM explanations of outcomes [15], as well as collective expert-driven fuzzy knowledge in compliance with data protection regulations [16].

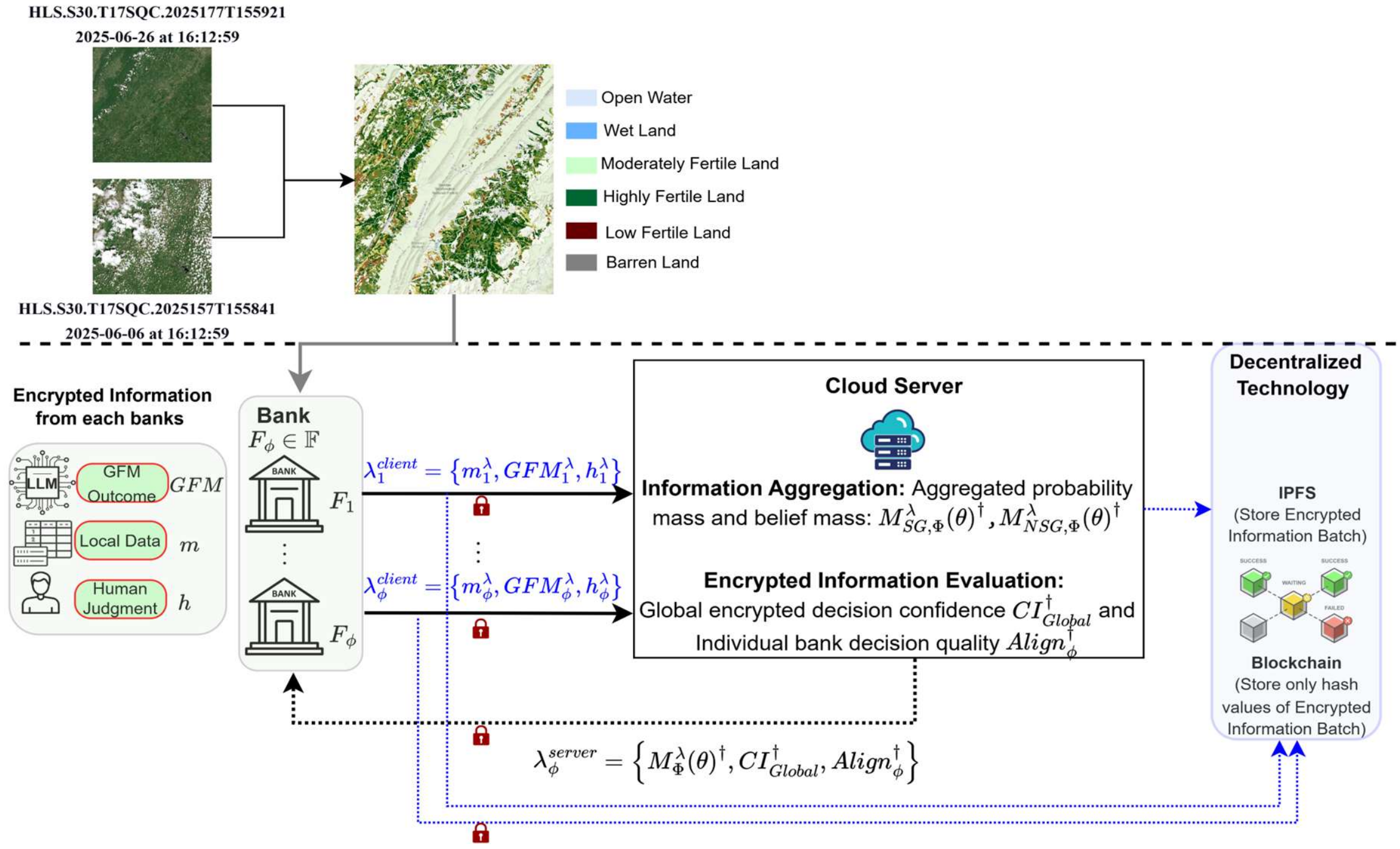

Figure 1. End-to-End Encrypted Multi-Bank Aggregation of Data-Driven Decisions, GFM Inference Outputs, and Human Expert-Derived Judgments

## III. Post-Quantum Secure Framework for Decentralized Multi-Bank Collaboration

The proposed post-quantum secure and blockchain-governed framework harnesses the federated intelligence of human experts and facilitates collaborative data aggregation among multiple financial institutions in a DeFi environment. Quantum-resilient security is achieved through the end-to-end encryption of data and transparent governance across participating banks, without centralized control or data exposure, as illustrated in Figure 1.

Let the consortium of participating banks be represented as $\mathbb{F} = \{F_1, \dots, F_\phi \dots, F_\Phi\}$. Human participants have two primary roles: (i) Delegators (domain experts) responsible for validating collaborative operations and auditing DeFi activities, such as verifying aggregated insights and model outcomes, and (ii) Delegates (operational users, such as data scientists or developers) to manage encrypted workflows and orchestrate on-chain data processes for DeFi applications.

The collaborative process is organized into discrete batches indexed by $\lambda \in \{1, \dots, \Lambda\}$. For a given batch $\lambda$, each bank $F_\phi \in \mathbb{F}$ contributes an encrypted information package:

$$\lambda_\phi^{client} = \{m_\phi^\lambda, LLM_\phi^\lambda, h_\phi^\lambda\} \quad (2)$$

Here, the encrypted information derived from a bank's local operations: the data-driven probability mass $m_\phi^\lambda$, the AI-generated decision $LLM_\phi^\lambda$ produced by a local GFM, and the human expert's belief assignment $h_\phi^\lambda$ to capture qualitative judgment and contextual reasoning.

### A. Setup for Lattice-Based Encrypted Computation

The local datasets of a bank $d_\phi$ are processed into an encrypted information batch comprising real numbers. At banks, it is first encoded into the polynomial ring and then encrypted using an RLWE or TRLE-based FHE scheme to enable privacy-preserving evaluation of mathematical functions directly over ciphertexts in a trustless server environment.

Each ciphertext is represented within the cyclotomic ring $\mathcal{R} = \mathbb{Z}_{q,N}[X]$ for RLWE and the torus cyclotomic ring

$\mathcal{R} = \mathbb{T}_N[N]$ for TRLWE. The parameters $N$ and $q$ govern the polynomial degree and modulus size, respectively. The binary secret key polynomials are sampled from $\mathbb{B}_{q,N}[X] = \frac{\mathbb{B}[X]}{(X^N+1)}$, where $\mathbb{B} \in \{0,1\}$. The security parameter $k$ determines the dimension of the secret key space for hardness against lattice attacks.

*A.1 Encode Batch of Financial Information*: Let a data point be denoted as $d \in [d_{min}, d_{max}] \subset \mathbb{R}$. The locally processed financial information and decision variables are encoded into polynomials $\psi_d(X) \in \mathcal{R}$ using the fixed precision encoding scheme:

$$\psi_d^{RLWE}: Enc(d) = \left\lfloor \frac{q}{2^{n_{msg}+n_{pad}}} \left\lfloor (d - d_{min}) \frac{2^{n_{msg}}}{\delta} \right\rceil \right\rceil \quad (3.1)$$

$$\psi_d^{TRLWE}: Enc(d) = \frac{\left\lfloor (d-d_{min}) \frac{2^{n_{msg}}}{\delta} \right\rceil}{2^{n_{msg}+n_{pad}}} \ (mod\ 1) \quad (3.2)$$

where $\delta = d_{max} - d_{min} + \sigma$ accounts for the interval range and noise tolerance, $n_{msg}$ is the message precision, $\sigma = \frac{d_{max}-d_{min}}{2^{n_{msg}}-1}$ bounds the quantization error, and $n_{pad}$ adds padding bits to absorb noise during encrypted arithmetic.

*A.2 Encryption and Decryption by Secret Key and Collective Public Key*: The vector of secret keys $s_\phi = \left(s_{\phi,1}(X), \ldots, s_{\phi,k}(X)\right) \in (\mathbb{B}_N[X])^k$, where each $s_{\phi,j}$ is a binary polynomial with coefficients in $\{0,1\}$ is sampled independently by each bank $F_\phi$.

A shared public vector $p_1(X) = \left(p_{1,1}(X), \ldots, p_{1,k}(X)\right)$is uniformly sampled from $\mathbb{Z}_{q,N}[X]$ or $\mathbb{T}_N[N]$, and agreed upon by all banks. Each bank $F_\phi$ computes its local public scalar $p_{0,\phi}(X)$ using a dot product over its vector secret:

$$p_{0,\phi}(X) = -\sum_{j=1}^{k} s_{\phi,j}(X) \cdot p_{1,j}(X) + \varepsilon_\phi(X)(mod\ q, X^N + 1) \quad (4)$$

where $\varepsilon_\phi(X) \sim \chi_\sigma$is local Gaussian noise with small coefficients to mask the secret and $q = 1$ for TRLWE. A bank only broadcasts the partial term $p_{0,\phi}(X)$ and adds local noise. A shared key vault stores and aggregates public keys, coordinated by delegates in banks:

$$p_0(X) = \sum_{\phi=1}^{\Phi} p_{0,\phi}(X) \quad (5)$$

The collective public key is $cpk = (p_0(X), p_1(X))$. It is used to encrypt a locally encoded data point $d$. A ciphertext of $d$ is:

$$\dagger_d = (p_0'(X), p_1(X)) \quad (6)$$

where $p_0'(X) = p_0(X) + \psi_d(X) + \varepsilon_\phi'(X)\ \ (mod\ q, X^N + 1)$ and $\varepsilon_\phi'(X)$ is an independent small noise polynomial satisfying $\varepsilon_\phi'(X) \neq \varepsilon_\phi(X)$. An example of a 32-bit FHE ciphertext is shown in Figure 2.

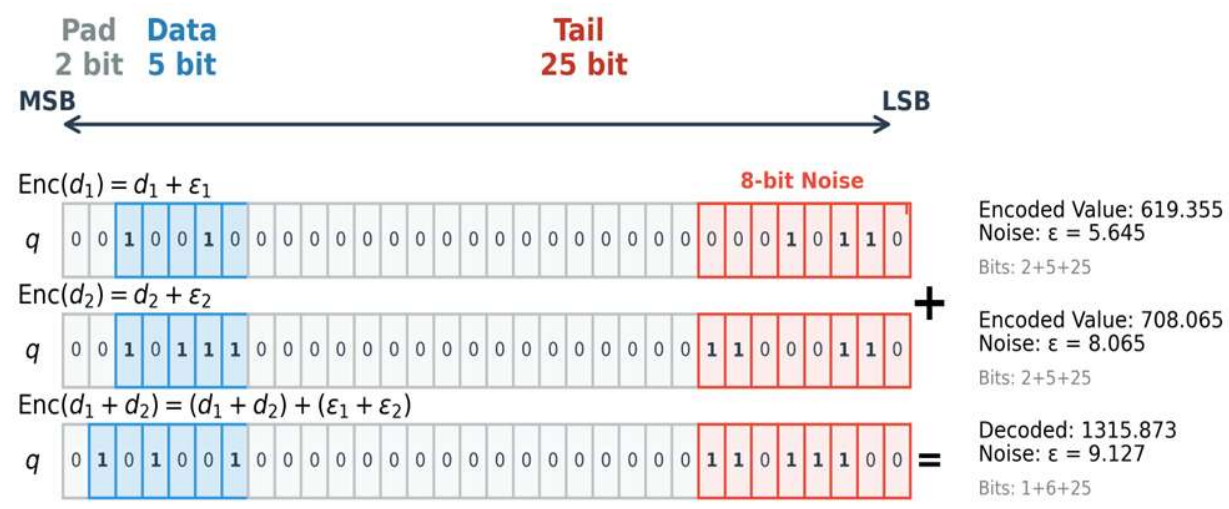

Figure 2. Addition (+) operation on a 32-bit sequence of FHE Ciphertexts

*A.3 Computation Infrastructure*: Table 1 summarizes the hybrid on-chain/off-chain architecture components required for end-to-end encrypted communication between multiple clients (multi-banks) and a server. It supports homomorphic computation for security, data confidentiality through cryptographic commitments, and transparent auditability.

Cloud infrastructure supports secure key storage and scalable compute resources required for FHE operations. The IPFS serves as a decentralized, content-addressed storage layer, where each bank's individual encrypted contribution $\lambda_\Phi^{client}$ and the cloud server's subsequent aggregated encrypted results $\lambda_\Phi^{Server}$ are stored. Consortium members deploy full blockchain nodes to record and verify the integrity of encrypted computations by storing $hash(\lambda_\Phi^{client})$ and $hash(\lambda_\Phi^{Server})$. The hash of the aggregated encrypted result is first computed and cross-verified against the individual hashes retrieved from IPFS; if validated, it is then recorded on-chain. Smart contracts autonomously manage the collective FHE key registry and maintain a tamper-proof audit log of computation events.

TABLE I. FHE-ENCRYPTED COMPUTATION INFRASTRUCTURE

| | |
|---|---|
| Off-chain Storage (Cloud) | *Cloud (Self-Hosted by each Bank)*: Raw financial data, metadata related to pre-trained GFM decisions, and human judgments |
| | *IPFS (Self-Hosted by each Bank)*: Distributed encrypted data storage |
| | *Shared Key Storage*: Collective public keys<br>*Private Key Storage:* Blockchain node identifiers and secret keys |
| On-Chain Storage (Blockchain) | *Blockchain:* Stores hashes of encrypted computations and hash of the aggregated encrypted result.<br>*Smart Contracts*: Manages key registry and computation logging |
| Computation (Cloud & Blockchain) | *Cloud FHE Servers:* Executes homomorphic operations (encrypted aggregation, evaluation) directly on ciphertexts.<br>*Smart Contract:* Coordinates audit trails and triggers verification routines. |

*B. Encrypted Computation between Clients and a Server*

Multiple financial institutions collaboratively contribute encrypted batch of information to enable privacy-preserving collective financial decision-making for groups of individuals underserved by local banks. The collaborative framework enriches their financial representations iteratively, where batch transitions from fundamental financial attributes to more complex attribute. A batch represents financial characteristics of similar subset of individuals.

For each iteration $i$, all participating bank $F_\phi$ submits a batch $\lambda_{\phi,i}^{client}$ containing three types information: (i) Basic Probability Mass (BPM) $m^\lambda = \left(m_1^\lambda, \dots, m_\phi^\lambda, \dots, m_\Phi^\lambda\right)$ derived from local data which quantifies belief distributions over potential financial outcomes, such as financial lending decisions, (ii) GFM-Generated Probabilistic Assessments $LLM^\lambda = \left(GFM_1^\lambda, \dots, GFM_\phi^\lambda, \dots, GFM_\Phi^\lambda\right)$ as AI reasoning, such as image classification, and (iii) Expert-Derived Belief Mass $h^\lambda = \left(h_1^\lambda, \dots, h_\phi^\lambda, \dots, h_\Phi^\lambda\right)$ to encode domain knowledge and contextual judgment.

*B.1 Federated Batch Selection*: The batch selection is conducted by an exploration–exploitation strategy. The exploration identifies borrower groups with high epistemic uncertainty (high entropy and distributed belief mass), whereas the exploitation prioritizes confident borrower groups (high singleton mass). Epistemic uncertainty is represented by probability mass over the powerset of the frame of discernment $\Theta = \{A, R, C\}$, where $A$ is approval, $R$ is rejection, and $C$ is conditional approval for a financial outcome. The powerset $2^\Theta \rightarrow [0,1]$ is:

$$2^\Theta = \{\phi, \{A\}, \{R\}, \{C\}, \{A, R\}, \{A, C\}, \{R, C\}, \Theta\} \quad (7.1)$$

Each batch's Basic Probability Assignment (BPA) satisfies:

$$\sum_{\theta \in 2^\Theta} m^\lambda(\theta) \leq 1 \quad (7.2)$$

A Unified Exploration–Exploitation Score $(\mathcal{U})$ for batch selection based on singleton subset concentration $m_{SG}^\lambda(\theta)$ (mass in definitive outcomes), non-singleton subset uncertainty $m_{NSG}^\lambda(\theta)$ (mass in ambiguous outcomes), and DST Entropy $m_{NSG}^\lambda(\theta)$ (measuring BPM dispersion) is:

$$\mathcal{U}\left(m^\lambda\right) = \alpha_1 m_{NSG}^\lambda(\theta) + \alpha_2 m_{SG}^\lambda(\theta) + \alpha_3 \frac{H(m^\lambda)}{H_{max}} \quad (8.1)$$

$$H\left(m^\lambda\right) = -\sum_{\theta \in 2^\Theta} m^\lambda(\theta) \log_2 m^\lambda(\theta) \quad (8.2)$$

where, $H_{max} = \log_2 8 = 3$. The parameters $\alpha_1, \alpha_2, \alpha_3 \in [0,1]$ satisfy $\alpha_1 + \alpha_2 + \alpha_3 = 1$, control the score dynamics such that $\alpha_1 > \alpha_2$ reflects exploration bias for uncertain cases, $\alpha_2 > \alpha_1$ reflects exploitation bias for confident cases, and higher $\alpha_3$ indicate increased sensitivity to entropy and distributional uncertainty. The boundary condition for maximum exploration and maximum exploitation is:

$$\lim_{\substack{m_{NSG}^\lambda \to 1 \\ H(m^\lambda) \to H_{max}}} \mathcal{U}\left(m^\lambda\right) = \alpha_1 + \alpha_3, \quad \lim_{\substack{m_{SG}^\lambda \to 1 \\ H(m^\lambda) \to 0}} \mathcal{U}\left(m^\lambda\right) = \alpha_2 \quad (9)$$

Batches with $\mathcal{U}\left(m^\lambda\right) \geq \theta_{THR}$ and $\mathcal{U}\left(m^\lambda\right) \leq \theta_{THR}$ are selected as exploration and exploitation set, respectively.

*B.2 Encrypted Computation at Server*: Each participating institution transmits an encrypted information batch $\lambda_{\phi,i}^{client}$ to server for homomorphic aggregation using DST. Both the BPM derived from local data and the belief masses assigned by in-house financial experts contain: (i) Singleton subsets $(SG)$, representing definitive outcomes $\theta \subset 2^\Theta: \{\{A\}, \{R\}, \{C\}\}$; (ii) Non-singleton subsets $(NSG)$, representing uncertainty $\theta \subset 2^\Theta: \{\{A, R\}, \{A, C\}, \{R, C\}, \Theta\}$; and (iii) Residual mass, representing ignorance, where the total mass in the power set is less than one.

*(i) Encrypted Information Aggregation*

Let, $\oplus$, $\ominus$, $\otimes$, and $o2$ represents homomorphic addition, subtraction, multiplication, and element-wise homomorphic squaring. The symbol † denotes encrypted computation.

The aggregated singleton mass is:

$$m_{SG,\Phi}^\lambda(\theta)^\dagger = \bigoplus_{\phi=1}^{\Phi} m_{SG,\phi}^\lambda(\theta)^\dagger \quad (10.1)$$

Similarly, the aggregated non-singleton mass is:

$$m_{NSG,\Phi}^\lambda(\theta)^\dagger = \bigoplus_{\phi=1}^{\Phi} m_{NSG,\phi}^\lambda(\theta)^\dagger \quad (10.2)$$

The remaining residual mass is computed as:

$$\widetilde{m}_{NSG,\Phi}^\lambda(\theta)^\dagger = 1^\dagger \ominus \left[m_{SG,\Phi}^\lambda(\theta)\right]^\dagger \ominus \left[m_{NSG,\Phi}^\lambda(\theta)\right]^\dagger \quad (11)$$

Here, $1^\dagger$ represents the encrypted unit constant. The normalized denominator for mass conservation is computed homomorphically as:

$$\Delta^\dagger = 1^\dagger \ominus \left[m_{NSG,\Phi}^\lambda(\theta)\right]^\dagger \quad (12.1)$$

The encrypted reciprocal is then approximated using Newton–Raphson iteration:

$$inv_\Delta{}^\dagger \cong \left[\frac{1}{\Delta}\right]^\dagger \quad (12.2)$$

Here, each iteration involves two homomorphic multiplications and one subtraction to achieve quadratic convergence for rapid approximation. The definitive financial outcome obtained by aggregation of singleton probability mass in an information batch by all participating institutions is:

$$\mathcal{M}_{SG,\Phi}^\lambda(\theta)^\dagger = \left[m_{SG,\Phi}^\lambda(\theta)\right]^\dagger \otimes \left[inv_\Delta\right]^\dagger \quad (13.1)$$

Similarly, the collective uncertainty or ignorance corresponding to non-singleton sets is:

$$\mathcal{M}_{NSG,\Phi}^{\lambda}(\theta)^{\dagger} = \left[\tilde{m}_{NSG,\Phi}^{\lambda}(\theta)\right]^{\dagger} \otimes \left[inv_{\Delta}\right]^{\dagger} \tag{13.2}$$

*(ii) Confidence Index*

The global encrypted confidence index $(CI_{Global})$ quantifies the dominance and uncertainty of the collective decision, expressed as:

$$CI_{Global}{}^{\dagger} = 1^{\dagger} \ominus \left[\mathcal{M}_{NSG,\Phi}^{\lambda}(\theta)\right]^{\dagger} \ominus \left(\bigoplus_{j \in \Theta}\left[\mathcal{M}_{SG,\Phi}^{\lambda}(j)\right]^{\dagger} \ominus max\left(\left[\mathcal{M}_{SG,\Phi}^{\lambda}(j)\right]^{\dagger}\right)^{o2}\right) \tag{14}$$

The alignment $(Align_{\phi})$ between a bank's singleton beliefs and the collective singleton beliefs is computed as:

$$m_{diff}{}^{\dagger} = \bigoplus_{j \in \Theta}\left[\left(m_{SG,\phi}^{\lambda}(j) - \mathcal{M}_{SG,\Phi}^{\lambda}(j)\right)^{o2}\right]^{\dagger} \tag{15.1}$$

$$Align_{\phi}{}^{\dagger} = 1^{\dagger} \ominus \left(0.5^{\dagger} \otimes m_{diff}{}^{\dagger}\right) \tag{15.2}$$

All operations are performed directly on ciphertexts; plaintext on aggregated probability and belief mass are not revealed to the server. The server returns the aggregate information $\lambda_{\Phi}^{Server}$, which is decrypted by the client secret key after conducting the collective key-switching process.

$$\lambda_{\Phi}^{Server} = \left\{\mathcal{M}_{\Phi}^{\lambda}(\theta)^{\dagger}, CI_{Global}{}^{\dagger}, Align_{\phi}{}^{\dagger}\right\} \tag{16}$$

Here, $\mathcal{M}_{\Phi}^{\lambda}(\theta)^{\dagger} = \left\{\mathcal{M}_{SG,\Phi}^{\lambda}(\theta)^{\dagger}, \mathcal{M}_{NSG,\Phi}^{\lambda}(\theta)^{\dagger}\right\}$.

## IV. Study on Rural Agricultural Lending

The proposed post-quantum DeFi for federated intelligence was empirically validated in a sandbox environment to enhance financial inclusion for agricultural lending in rural Virginia, USA. A consortium of four regional banks serving northern (Rockingham, Augusta, Orange counties) and southern (Pittsylvania, Halifax, Mecklenburg counties) rural communities collaboratively participated in the end-to-end encrypted collective financial decision-making process for borrowers categorized by delinquency risk level: {*low, medium, high*} and lending decisions were classified as *{A (approval), R (rejection), C (conditional approval)}*. Land-fertility conditions were validated using pre-trained IBM–NASA Prithvi geospatial models applied to NASA Harmonized Landsat–Sentinel-2 (HLS v2.0) satellite imagery as *{moderate, high, low, open water, wet land}*.

All inter-bank data exchange and server-side aggregation were performed under end-to-end homomorphic encryption using multi-party RLWE and TRLWE schemes configured for 128-bit post-quantum security (NIST Level 1 equivalent). Cryptographic parameters are summarized in Table II. The homomorphic computation in the server was hosted on an Azure Virtual Machine with 16 vCPU and 32 GiB RAM.

The federated intelligence workflow was executed over three iterative rounds between the participating institutions and the aggregation server. Each iteration processed encrypted batches of 16 borrower risk profiles, representing groups of applicants with financial characteristics spanning from foundational metrics (credit scores, income stability) to higher-dimensional financial features (seasonal cash-flow patterns, commodity price exposure, and weather-related risk indices).

TABLE II. RLWE and TRLWE Parameters (Security 128-bit)

| Parameters | RLWE | TRLWE |
|---|---|---|
| Ring Dimension ($N$) | $N = 2^{12} = 4096$ | $N = 2^{10} = 1024$ |
| Secret Key Distribution | Ternary $\{-1, 0, +1\}$ | Binary $\{0,1\}$ |
| Error Distribution ($\sigma$) | 3.19 | $2^{-15}$ |
| Ciphertext Modulus | $Log\ Q = 218$ | 1 (torus $\mathbb{T}$) |
| Multiplication Depth | ≥20 levels with rescaling | Unlimited (Bootstrapping) |
| Average Ciphertext Size per Iteration (MB) | [0.00056, 0.00102, 0.00102] | [0.00102, 0.00303, 0.00304] |

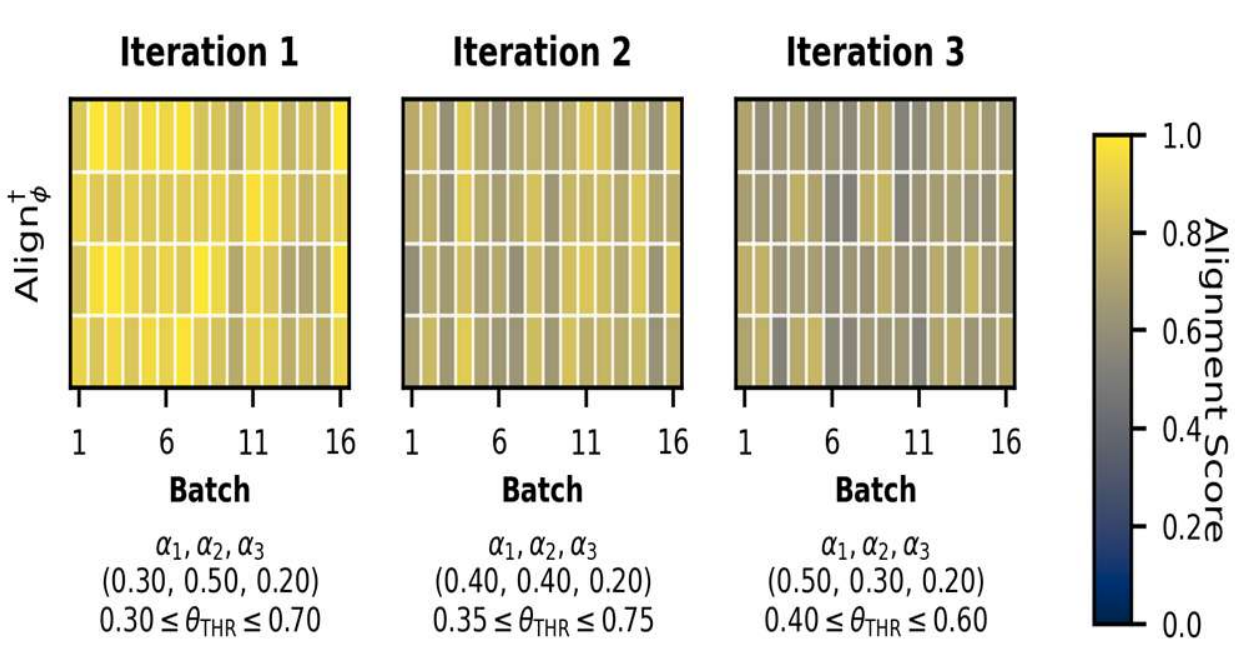

Figure 3. Selected batch parameters and inter-bank alignment scores, decrypted at the client after homomorphic computation on the server

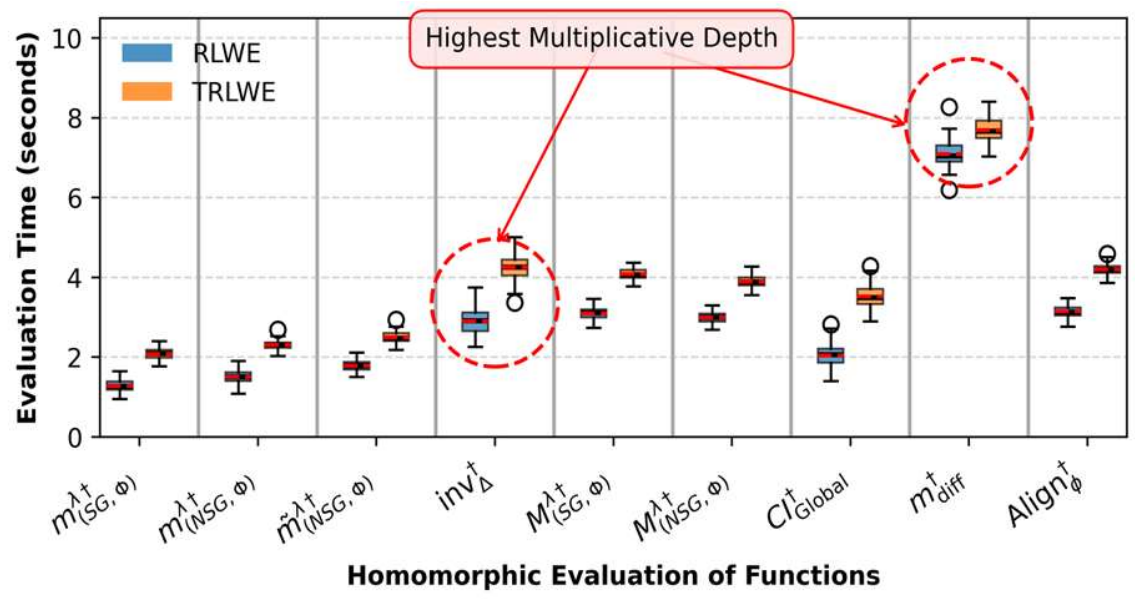

Figure 4. Homomorphic evaluation time for encrypted functions (Expression 9.1 to 14.2) in the server under RLWE and TRLWE schemes

The progression of batch selection across the three iterations transitioned from an exploitation-dominant strategy to establish stable baseline predictions, to a balanced approach to representational diversity, and finally to an exploration-dominant strategy to prioritize complex and uncertain cases. Figure 3 illustrates the batch-selection parameters and the alignment index ($Align_{\phi}$) of decisions derived from each bank's data and in-house expert assessments. It is computed homomorphically at the server to evaluate contribution quality. The distribution of homomorphic evaluation time for the multivariate functions (Expressions 9.1–14.2) executed at the server using RLWE- and TRLWE-based FHE schemes is shown in Figure 5. Among all functions, the encrypted reciprocal $inv_{\Delta}{}^{\dagger}$ and the alignment-difference $m_{diff}{}^{\dagger}$ computation exhibited the highest evaluation times in both schemes due to their greater multiplicative depth. The enhanced decision confidence for a selected group of underserved rural borrowers processed collaboratively in multiple batches is illustrated geographically on the Virginia map. The decision outcome distributions (A: approval, R: rejection, C: conditional) across three iterations show that TRLWE-based FHE achieves stronger confidence calibration and improved delinquency risk management relative to RLWE. However, performance remains slightly below unprocessed data used to retrain the AI model due to residual encrypted–decrypted noise inherent in homomorphic computation.

Both the pre-aggregated encrypted batches $\lambda_{\phi}^{client}$ from each bank and the server-aggregated results $\lambda_{\Phi}^{Server}$ are stored in IPFS for decentralized persistence (average upload time: 12 $ms$ per batch; CID retrieval: 8 $ms$). The corresponding SHA-256 content hashes are immutably recorded on the Polygon L2 blockchain (average confirmation time: 2.1 $s$; gas usage: 18k; cost: \$0.003) for tamper-evident, fully auditable recordkeeping, and time-stamped provenance for every encrypted contribution and aggregation outcome.

## V. Conclusion

The proposed post-quantum federated DeFi framework addresses emerging security requirements through lattice-based FHE for privacy-preserving multi-bank decision intelligence by encrypted aggregation of human–AI belief fusion. It supports auditable inter-bank financial decisions and enhances inclusivity for underserved borrowers.

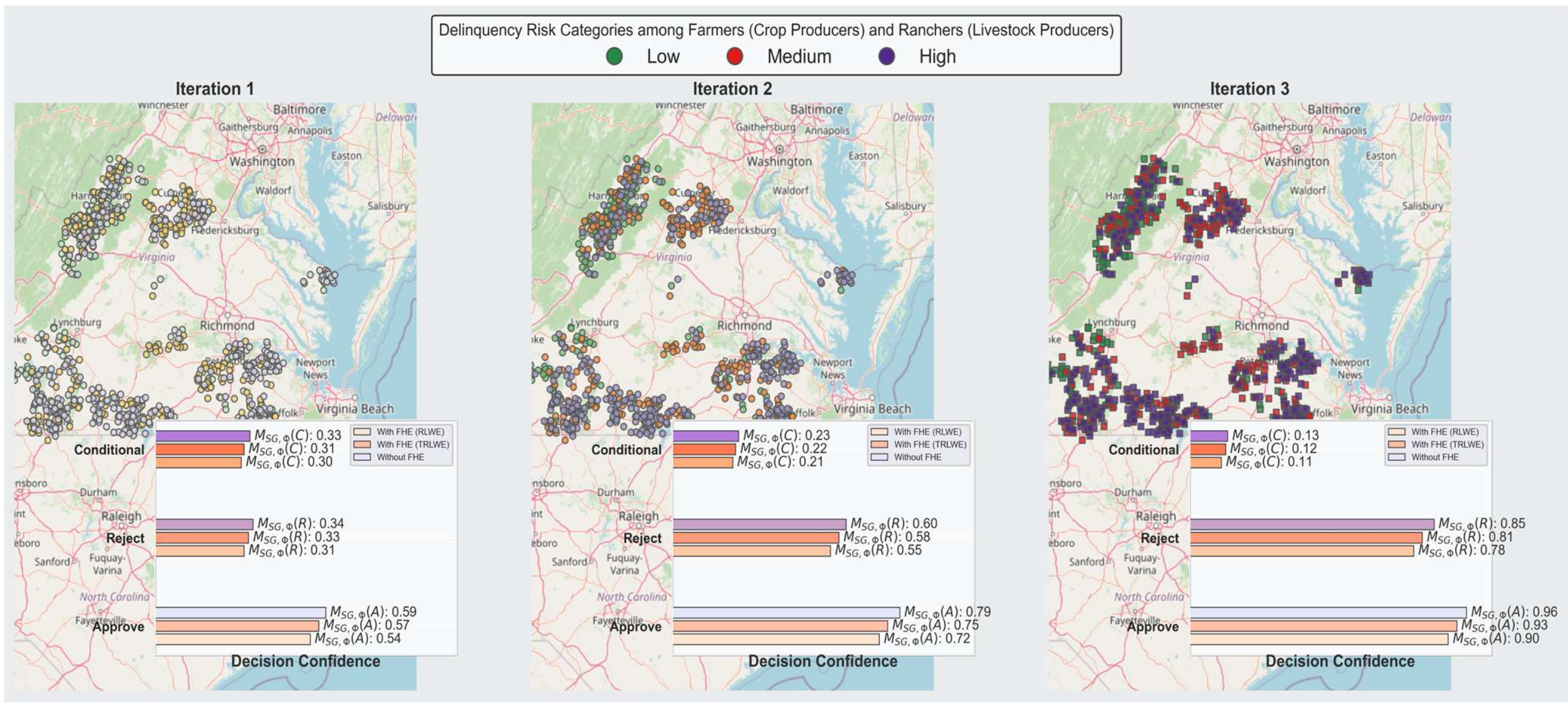

Figure 5. Iterative improvement in decision confidence and reduction of delinquency risk among farmers and ranchers in Virginia, USA.